# Harmonic State Space Modeling of a Three-Phase Modular Multilevel Converter

Jing Lyu[1], Marta Molinas[1], Xu Cai[2]

1. Department of Engineering Cybernetics, Norwegian University of Science and Technology, Trondheim 7491, Norway
2. Wind Power Research Center, Shanghai Jiao Tong University, Shanghai 200240, China

***Abstract*—This paper presents the harmonic state space (HSS) modeling of a three-phase modular multilevel converter (MMC). MMC is a converter system with a typical multi-frequency response due to its significant harmonics in the arm currents, capacitor voltages, and control signals. These internal harmonic dynamics can have a great influence on the operation characteristics of MMC. However, the conventional modeling methods commonly used in two-level voltage-source converters (VSCs), where only the fundamental-frequency dynamic is considered, will lead to an inaccurate model that cannot accurately reflect the real dynamic characteristics of MMC. Therefore, the HSS modeling method, in which harmonics of state variables, inputs, and outputs are posed separately in a state-space form, is introduced in this paper to model the MMC in order to capture all the harmonics and the frequency couplings. The steady-state and small-signal dynamic HSS models of a three-phase MMC are developed, respectively. The validity of the developed HSS model of a three-phase MMC has been verified by the results from both the nonlinear time domain simulation model in MATLAB/Simulink and the laboratory prototype with 12 submodules per arm.**

## I. INTRODUCTION

In recent years, modular multilevel converter (MMC) has already been widely used in high-voltage/high-power applications, e.g., high-voltage direct current (HVDC) transmission, due to its advantages, such as modular design, low switching loss, low distortion of output voltage, easily scalable in terms of voltage levels, and so on [1], [2]. However, because of its complex internal structure, the modeling and control of MMC is much more complicated than that of two-level voltage-source converters (VSCs) [3], [4]. In addition, the internal dynamics of MMC such as capacitor voltage fluctuations and harmonic circulating currents may have detrimental effects on the stable operation of MMC interconnected power systems, especially for applications of renewable energy integration [5], [6]. Therefore, it is essential to consider the internal dynamics of MMC when concerning the harmonic interaction and small-signal stability issues.

In the beginning, most of the work on the modeling of MMC focused on the external characteristics of MMC [7]-[9], where the internal dynamics were completely neglected, resulting in essentially the same models as two-level VSCs. These models truly facilitate the large power system studies on condition that enough internal damping is guaranteed in the MMC itself. However, MMC is widely used in high-voltage applications, where weak internal damping is a common phenomenon. In this case, the instability issues caused by the internal dynamics of MMC cannot be accurately identified by using the previous models. More recently, several researchers have made efforts on the modeling of MMC considering its internal dynamics [10]-[16]. Dragan *et al.* [10], [11] developed the DQ model of MMC in multiple DQ rotating coordinate frames, including dc, fundamental frequency, and second harmonic, and based on this idea, Ref. [12] and [13] considered third harmonic. However, this modeling method involves lengthy algebra and is difficult to be extended to high number of harmonics. Ref. [14]-[16] derived the MMC models in the time periodic framework, which have high accuracy, but have the same shortcomings as Dragan's modeling method.

In order to accurately model the MMC and to readily extend to high number of harmonics for harmonic interaction studies, the harmonic state space (HSS) modeling method, which is able to build the multi-harmonic model of power converters, is first introduced in this paper to model a three-phase MMC. The HSS model, in which harmonics of state variables, inputs, and outputs are posed separately in a state-space form, can be used for dynamic small-signal studies and for steady-state harmonic interaction studies. A linear time periodic (LTP) system in time domain can be transformed into a linear time invariant (LTI) system in frequency domain by the HSS modeling method. The HSS modeling method has been used to model the linear time periodic (LTP) system in many fields of power systems since the 1990s [17]-[22], for instance, buck-boost converter, thyristor-controlled reactor (TCR), and two-level VSCs. However, the HSS modeling of MMC has not so far been reported in the literature. This paper will present the HSS modeling of a three-phase MMC. The steady state and small-signal dynamic HSS models of a three-phase MMC are developed, respectively. The validity of the developed HSS models of the three-phase MMC is verified by the results from both the nonlinear time domain simulation in MATLAB/Simulink and the laboratory setup with 12 submodules per arm.

The rest of the paper is organized as follows. Section II describes the formulation of the HSS modeling. Section III presents the steady state HSS modeling of a three-phase MMC, and nonlinear time domain simulation results for validation of the developed steady state HSS modeling of the three-phase MMC. The small-signal dynamic HSS modeling of the three-phase MMC is shown in Section IV, and the time domain simulation results are also given in this section to verify the small-signal dynamic HSS model. Section V demonstrates the experimental results to further validate the developed HSS model of the three-phase MMC. Section VI concludes the paper.

## II. FORMULATION OF HARMONIC STATE SPACE (HSS) MODELING

For any time varying periodic signal $x(t)$, it can be written in the form of Fourier series as:

$$x(t) = \sum_{k\in\mathbb{Z}} X_k e^{jk\omega_1 t} \tag{1}$$

where $\omega_1=2\pi/T$, $T$ is the fundamental period of the signal, and $X_k$ is the Fourier coefficient that can be calculated by

$$X_k = \frac{1}{T}\int_{t_0}^{t_0+T} x(t) e^{-jk\omega_1 t} dt \tag{2}$$

The state space equation of a linear time periodic (LTP) system can be expressed as

$$\dot{x}(t) = A(t)x(t) + B(t)u(t) \tag{3}$$

Based on the Fourier series and harmonic balance theory, the state space equation in time domain can be transformed into the harmonic state-space equation in frequency domain, which is like

$$s\mathrm{X} = (\mathrm{A} - \mathrm{Q})\mathrm{X} + \mathrm{BU} \tag{4}$$

where X, U, A, B, and N are indicated as (5)~(9), respectively, of which the elements $X_h$, $U_h$, $A_h$, and $B_h$ are the Fourier coefficients of the $h$th harmonic of $x(t)$, $u(t)$, $A(t)$, and $B(t)$ in (3), respectively. Note that A and B are Toeplitz matrices in order to perform the frequency domain convolution operation, Q is a diagonal matrix that represents the frequency information, and $I$ is an identity matrix. In addition, it should be noted that the harmonic order $h$ considered in the HSS model should be selected according to the requirements for accuracy and complexity of the model.

$$\mathrm{X}=\left[X_{-h},\cdots,X_{-1},X_0,X_1,\cdots,X_h\right]^T \tag{5}$$

$$\mathrm{U}=\left[U_{-h},\cdots,U_{-1},U_0,U_1,\cdots,U_h\right]^T \tag{6}$$

$$\mathrm{A}=\begin{bmatrix} A_0 & A_{-1} & \cdots & A_{-h} & & & \\ A_1 & \ddots & \ddots & \ddots & \ddots & & \\ \vdots & \ddots & A_0 & A_{-1} & \ddots & \ddots & \\ A_h & \ddots & A_1 & A_0 & A_{-1} & \ddots & A_{-h} \\ & \ddots & \ddots & A_1 & A_0 & \ddots & \vdots \\ & & \ddots & \ddots & \ddots & \ddots & A_{-1} \\ & & & A_h & \cdots & A_1 & A_0 \end{bmatrix} \tag{7}$$

$$\mathrm{B}=\begin{bmatrix} B_0 & B_{-1} & \cdots & B_{-h} & & & \\ B_1 & \ddots & \ddots & \ddots & \ddots & & \\ \vdots & \ddots & B_0 & B_{-1} & \ddots & \ddots & \\ B_h & \ddots & B_1 & B_0 & B_{-1} & \ddots & B_{-h} \\ & \ddots & \ddots & B_1 & B_0 & \ddots & \vdots \\ & & \ddots & \ddots & \ddots & \ddots & B_{-1} \\ & & & B_h & \cdots & B_1 & B_0 \end{bmatrix} \tag{8}$$

$$\mathrm{Q}=\mathrm{diag}\left[-jh\omega_1,\cdots,-j\omega_1,0,j\omega_1,\cdots,jh\omega_1\right] \tag{9}$$

## III. STEADY STATE HSS MODEL OF A THREE-PHASE MMC

Fig. 1 shows the circuit diagram of a three-phase MMC. Each phase-leg of the MMC consists of one upper and one lower arm connected in series between the dc terminals. Each arm consists of $N$ identical series-connected submodules (SMs), one arm inductor $L$, and an arm equivalent series resistor $R$. Each SM contains a half-bridge as a switching element and a dc storage capacitor $C_{SM}$. In high-voltage applications, $N$ may be as high as several hundreds. It should be noted that the SM may use a half-bridge or a full-bridge topology, which, however, doesn't affect the discussion here.

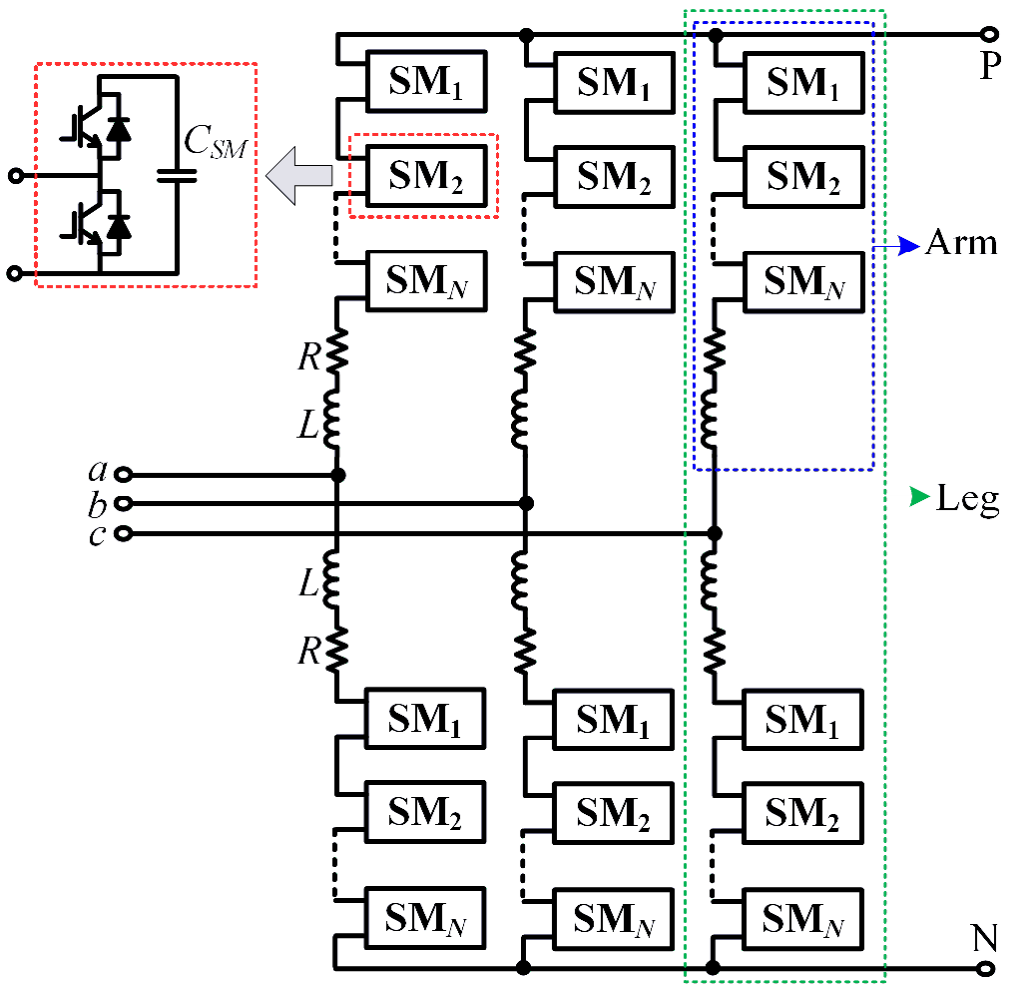

Fig. 1. Circuit diagram of a three-phase MMC.

Fig. 2 presents the average-value model of one phase leg of MMC, where $C_{arm}=C_{SM}/N$, $C_{SM}$ is the submodule capacitance, $L$ and $R$ are the arm inductance and equivalent series resistance, $i_u$ and $i_l$ are the upper and lower arm currents, $v_{cu}^{\Sigma}$ and $v_{cl}^{\Sigma}$ are the sum capacitor voltages of the upper and lower arms, $v_g$ and $i_g$ are the ac-side phase voltage and current, $i_c$ is the circulating current, $n_u$ and $n_l$ are the insertion indices of the upper and lower arms, and $V_{dc}$ is the dc bus voltage. In addition, $Z_L(=R_L+j\omega_1 L_L)$ is the ac-side equivalent load in order to determine the steady state operating point.

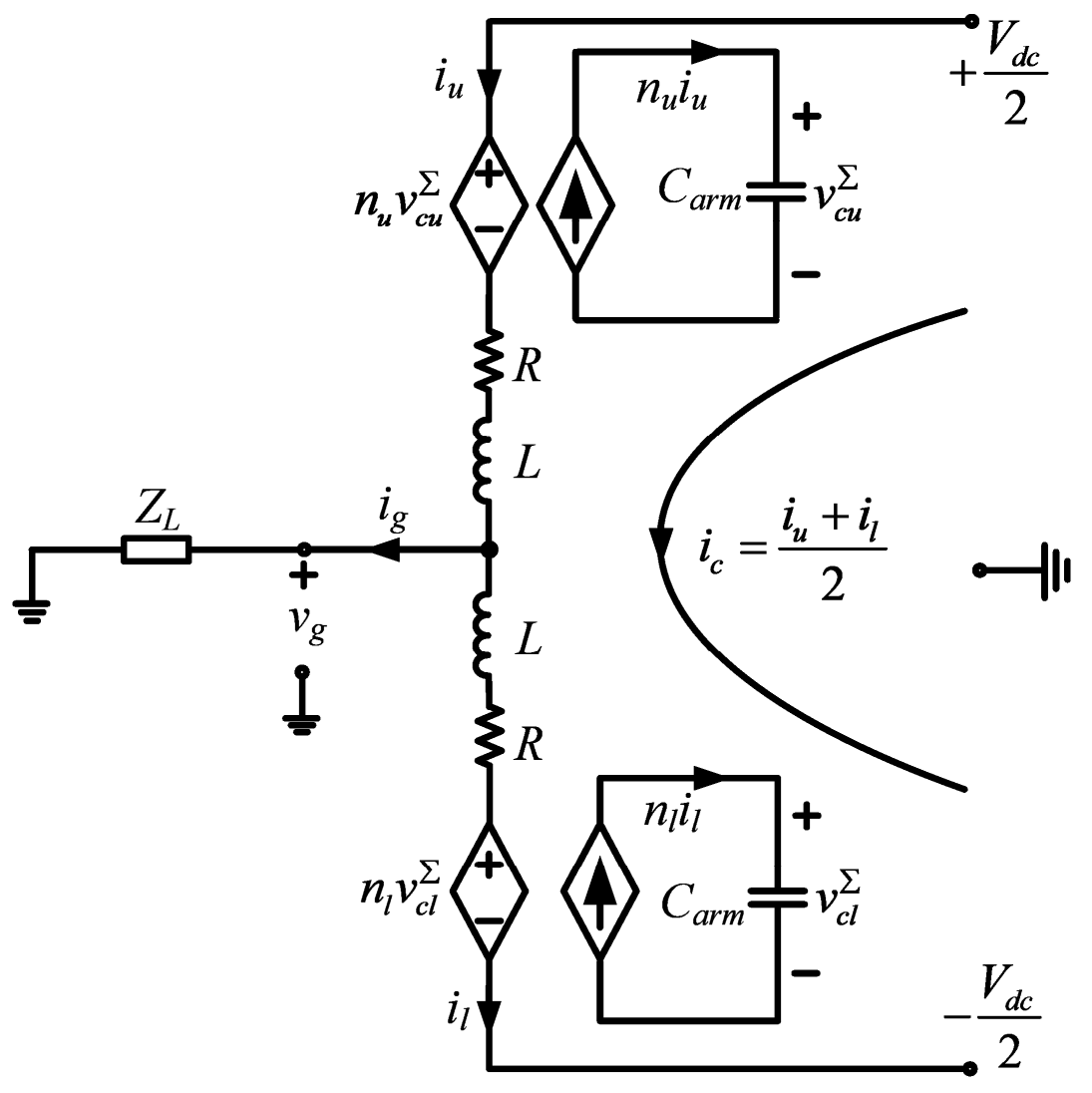

Fig. 2. Average-value model of one phase leg of MMC.

According to Fig. 2, one can obtain

$$\frac{di_{cx}}{dt} = -\frac{R}{L}i_{cx} - \frac{n_{ux}}{2L}v_{cux}^{\Sigma} - \frac{n_{lx}}{2L}v_{clx}^{\Sigma} + \frac{v_{dc}}{2L}, (x = a,b,c) \tag{10}$$

$$\frac{dv_{cux}^{\Sigma}}{dt} = \frac{n_{ux}}{C_{arm}}i_{cx} + \frac{n_{ux}}{2C_{arm}}i_{gx}, (x = a,b,c) \tag{11}$$

$$\frac{dv_{clx}^{\Sigma}}{dt} = \frac{n_{lx}}{C_{arm}}i_{cx} - \frac{n_{lx}}{2C_{arm}}i_{gx}, (x = a,b,c) \tag{12}$$

$$\frac{di_{gx}}{dt} = -\frac{n_{ux}}{L}v_{cux}^{\Sigma} + \frac{n_{lx}}{L}v_{clx}^{\Sigma} - \frac{R}{L}i_{gx} - \frac{2v_{gx}}{L}, (x = a,b,c) \tag{13}$$

$$v_{gx} = Z_L i_{gx}, (x = a,b,c) \tag{14}$$

The above differential equations can be expressed as a form of time domain state space equation in (3), where

$$x(t) = \left[i_{ca}, i_{cb}, i_{cc}, v_{cua}^{\Sigma}, v_{cub}^{\Sigma}, v_{cuc}^{\Sigma}, v_{cla}^{\Sigma}, v_{clb}^{\Sigma}, v_{clc}^{\Sigma}, i_{ga}, i_{gb}, i_{gc}\right]^T \tag{15}$$

$$u(t) = [v_{dc}]^T \tag{16}$$

$$A(t)=\begin{bmatrix} -\frac{R}{L} & 0 & 0 & -\frac{n_{ua}}{2L} & 0 & 0 & -\frac{n_{la}}{2L} & 0 & 0 & 0 & 0 & 0 \\ 0 & -\frac{R}{L} & 0 & 0 & -\frac{n_{ub}}{2L} & 0 & 0 & -\frac{n_{lb}}{2L} & 0 & 0 & 0 & 0 \\ 0 & 0 & -\frac{R}{L} & 0 & 0 & -\frac{n_{uc}}{2L} & 0 & 0 & -\frac{n_{lc}}{2L} & 0 & 0 & 0 \\ \frac{n_{ua}}{C_{arm}} & 0 & 0 & 0 & 0 & 0 & 0 & 0 & 0 & \frac{n_{ua}}{2C_{arm}} & 0 & 0 \\ 0 & \frac{n_{ub}}{C_{arm}} & 0 & 0 & 0 & 0 & 0 & 0 & 0 & 0 & \frac{n_{ub}}{2C_{arm}} & 0 \\ 0 & 0 & \frac{n_{uc}}{C_{arm}} & 0 & 0 & 0 & 0 & 0 & 0 & 0 & 0 & \frac{n_{uc}}{2C_{arm}} \\ \frac{n_{la}}{C_{arm}} & 0 & 0 & 0 & 0 & 0 & 0 & 0 & 0 & -\frac{n_{la}}{2C_{arm}} & 0 & 0 \\ 0 & \frac{n_{lb}}{C_{arm}} & 0 & 0 & 0 & 0 & 0 & 0 & 0 & 0 & -\frac{n_{lb}}{2C_{arm}} & 0 \\ 0 & 0 & \frac{n_{lc}}{C_{arm}} & 0 & 0 & 0 & 0 & 0 & 0 & 0 & 0 & -\frac{n_{lc}}{2C_{arm}} \\ 0 & 0 & 0 & -\frac{n_{ua}}{L} & 0 & 0 & \frac{n_{la}}{L} & 0 & 0 & -\frac{R+2Z_L}{L} & 0 & 0 \\ 0 & 0 & 0 & 0 & -\frac{n_{ub}}{L} & 0 & 0 & \frac{n_{lb}}{L} & 0 & 0 & -\frac{R+2Z_L}{L} & 0 \\ 0 & 0 & 0 & 0 & 0 & -\frac{n_{uc}}{L} & 0 & 0 & \frac{n_{lc}}{L} & 0 & 0 & -\frac{R+2Z_L}{L} \end{bmatrix} \tag{17}$$

$$B=\left[\frac{1}{2L},\frac{1}{2L},\frac{1}{2L},0,0,0,0,0,0,0,0,0\right]^T \tag{18}$$

The time domain state space model can be transformed into the frequency domain state space model by the HSS modeling method:

$$s\mathbf{X}=\mathbf{AX}+\mathbf{BU} \tag{19}$$

where

$$\mathbf{X}=\left[I_{ca},I_{cb},I_{cc},V_{cua}^{\Sigma},V_{cub}^{\Sigma},V_{cuc}^{\Sigma},V_{cla}^{\Sigma},V_{clb}^{\Sigma},V_{clc}^{\Sigma},I_{ga},I_{gb},I_{gc}\right]^T \tag{20}$$

$$\mathbf{U}=\left[V_{dc}\right]^T \tag{21}$$

$$\mathbf{A}=\begin{bmatrix} -\frac{R}{L}I-Q & O & O & -\frac{\Gamma[N_{ua}]}{2L} & O & O & -\frac{\Gamma[N_{la}]}{2L} & O & O & O & O & O \\ O & -\frac{R}{L}I-Q & O & O & -\frac{\Gamma[N_{ub}]}{2L} & O & O & -\frac{\Gamma[N_{lb}]}{2L} & O & O & O & O \\ O & O & -\frac{R}{L}I-Q & O & O & -\frac{\Gamma[N_{uc}]}{2L} & O & O & -\frac{\Gamma[N_{lc}]}{2L} & O & O & O \\ \frac{\Gamma[N_{ua}]}{C_{arm}} & O & O & -Q & O & O & O & O & O & \frac{\Gamma[N_{ua}]}{2C_{arm}} & O & O \\ O & \frac{\Gamma[N_{ub}]}{C_{arm}} & O & O & -Q & O & O & O & O & O & \frac{\Gamma[N_{ub}]}{2C_{arm}} & O \\ O & O & \frac{\Gamma[N_{uc}]}{C_{arm}} & O & O & -Q & O & O & O & O & O & \frac{\Gamma[N_{uc}]}{2C_{arm}} \\ \frac{\Gamma[N_{la}]}{C_{arm}} & O & O & O & O & O & -Q & O & O & -\frac{\Gamma[N_{la}]}{2C_{arm}} & O & O \\ O & \frac{\Gamma[N_{lb}]}{C_{arm}} & O & O & O & O & O & -Q & O & O & -\frac{\Gamma[N_{lb}]}{2C_{arm}} & O \\ O & O & \frac{\Gamma[N_{lc}]}{C_{arm}} & O & O & O & O & O & -Q & O & O & -\frac{\Gamma[N_{lc}]}{2C_{arm}} \\ O & O & O & -\frac{\Gamma[N_{ua}]}{L} & O & O & \frac{\Gamma[N_{la}]}{L} & O & O & -\frac{R+2Z_L}{L}I-Q & O & O \\ O & O & O & O & -\frac{\Gamma[N_{ub}]}{L} & O & O & \frac{\Gamma[N_{lb}]}{L} & O & O & -\frac{R+2Z_L}{L}I-Q & O \\ O & O & O & O & O & -\frac{\Gamma[N_{uc}]}{L} & O & O & \frac{\Gamma[N_{lc}]}{L} & O & O & -\frac{R+2Z_L}{L}I-Q \end{bmatrix} \tag{22}$$

$$B=\left[\frac{1}{2L}I,\frac{1}{2L}I,\frac{1}{2L}I,O,O,O,O,O,O,O,O,O\right]^{T} \quad (23)$$

in which $\Gamma[\ ]$ means the Toeplitz matrix of the Fourier coefficients of the time varying variables, $I$ is identity matrix, $O$ is zero matrix. Additionally, the lowercase letters in (15)~(17) denote the time domain signals, the uppercase letters in (20)~(22) represent the Fourier coefficients from [-$h$… -1,0,1… $h$]. Furthermore, the Toeplitz matrices of the Fourier coefficients of the insertion indices are given in Appendix (A1)~(A6) (taking $h$=3 for instance).

Letting the left side of (19) to be zero, the steady state harmonic components of the state variables can thus be calculated by

$$\mathbf{X}_{ss}=-\mathbf{A}^{-1}\mathbf{BU} \quad (24)$$

It is pointed out that the frequency domain signals obtained from the HSS model in (24) can be converted into the time domain signals by using (1), and these time domain signals can then be compared with those signals obtained from the nonlinear time domain simulation model in MATLAB/Simulink.

In order to verify the steady state HSS model of MMC developed in this paper, a comparison between the steady state results of the nonlinear time domain simulation model and the HSS model of MMC has been carried out, as shown in Fig. 3, where the nonlinear time domain simulation model is implemented in MATLAB/Simulink, and the HSS model is performed by using an m-file in MATLAB. In this case, open-loop control is employed in the MMC by directly setting the sinusoidal modulation voltages. The harmonic order considered in the HSS model is $h$=3. The main electrical parameters of the MMC in the simulation are listed as follows: rated power $P_N$ = 50 MW, nominal ac line RMS voltage $V_N$ = 166 kV, nominal dc-bus voltage $V_{dc}$ = 320 kV, fundamental angular frequency $\omega_1$ = 314 rad/s, SM number per arm $N$ = 20, SM capacitance $C_{SM}$ = 140 μF, arm inductance $L$ = 360 mH, and arm resistance $R$ = 1 Ω. In addition, pure resistive load is used in this case.

As can be seen, the results between the steady state HSS model and the time domain simulation model have a good match, which shows that the steady state HSS model of MMC developed in this paper is able to capture all the harmonics in the circulating currents and capacitor voltages, and is accurate enough for harmonic steady-state studies. In general, the circulating currents mainly contains dc and second harmonic components as well as other even harmonics which are negligibly small in normal operation, and the capacitor voltages theoretically contains all the harmonics, in which, however, the dc, fundamental, second and third harmonic components are dominant in normal cases. In addition, the ac phase currents can approximately be treated as harmonic-free.

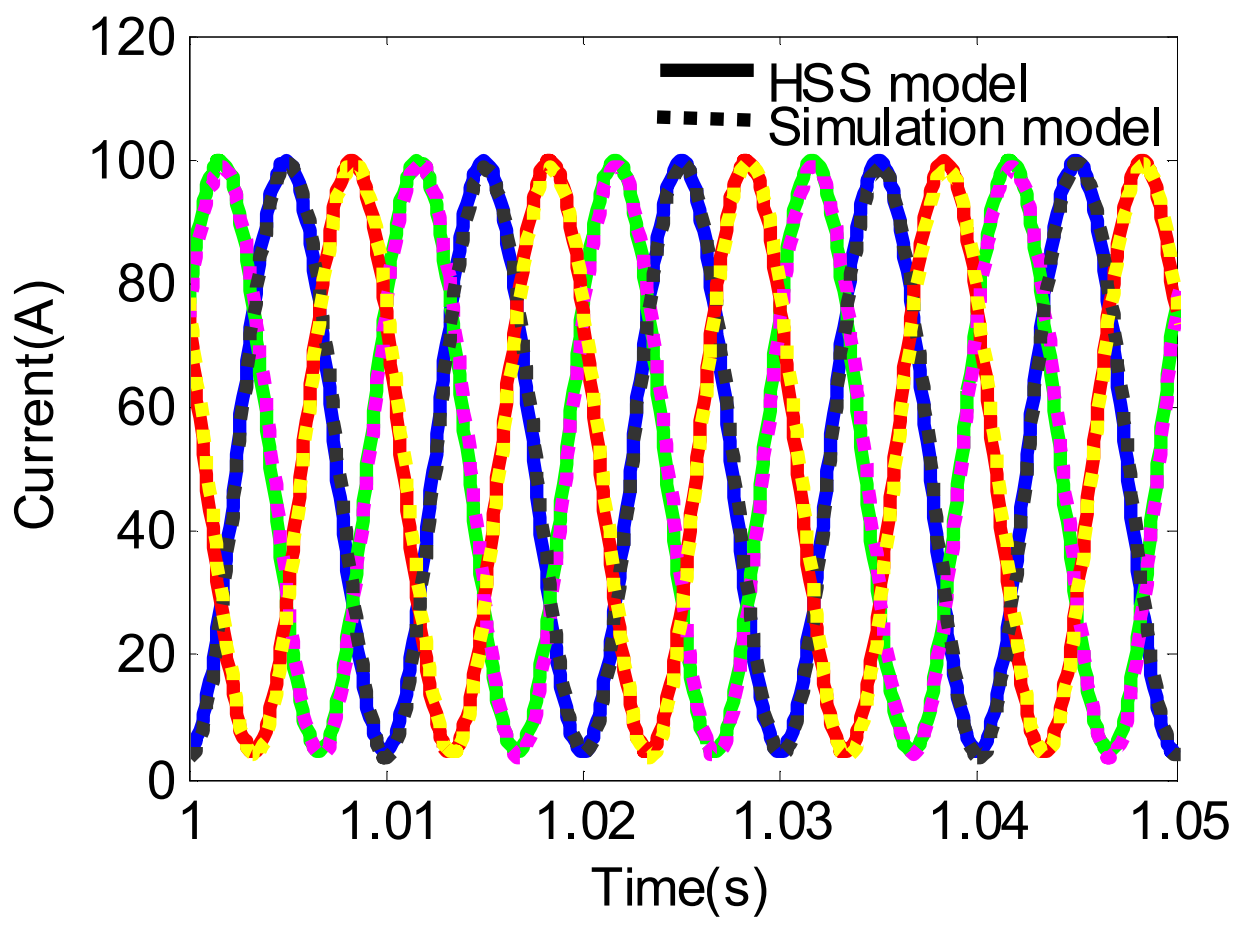

(a)

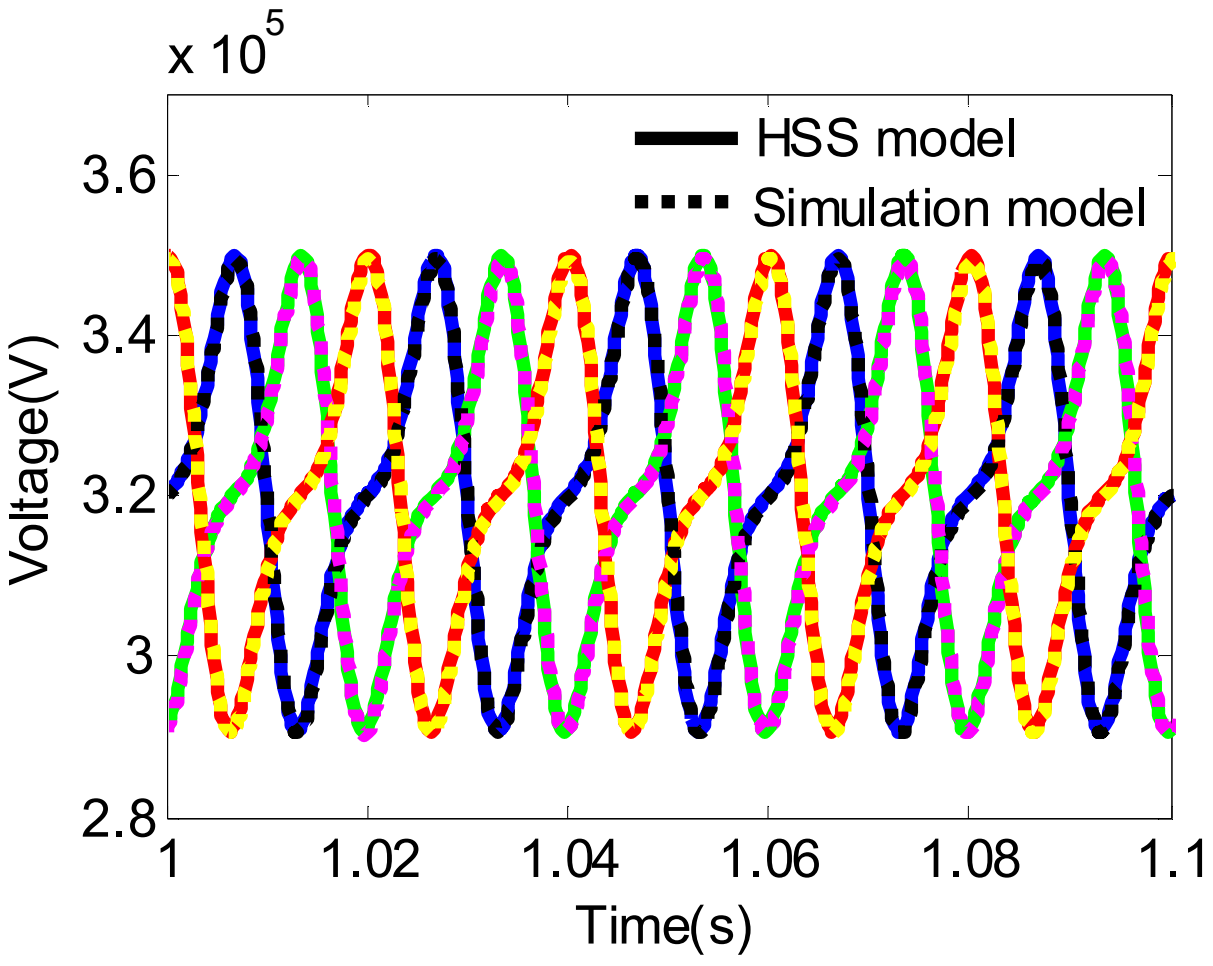

(b)

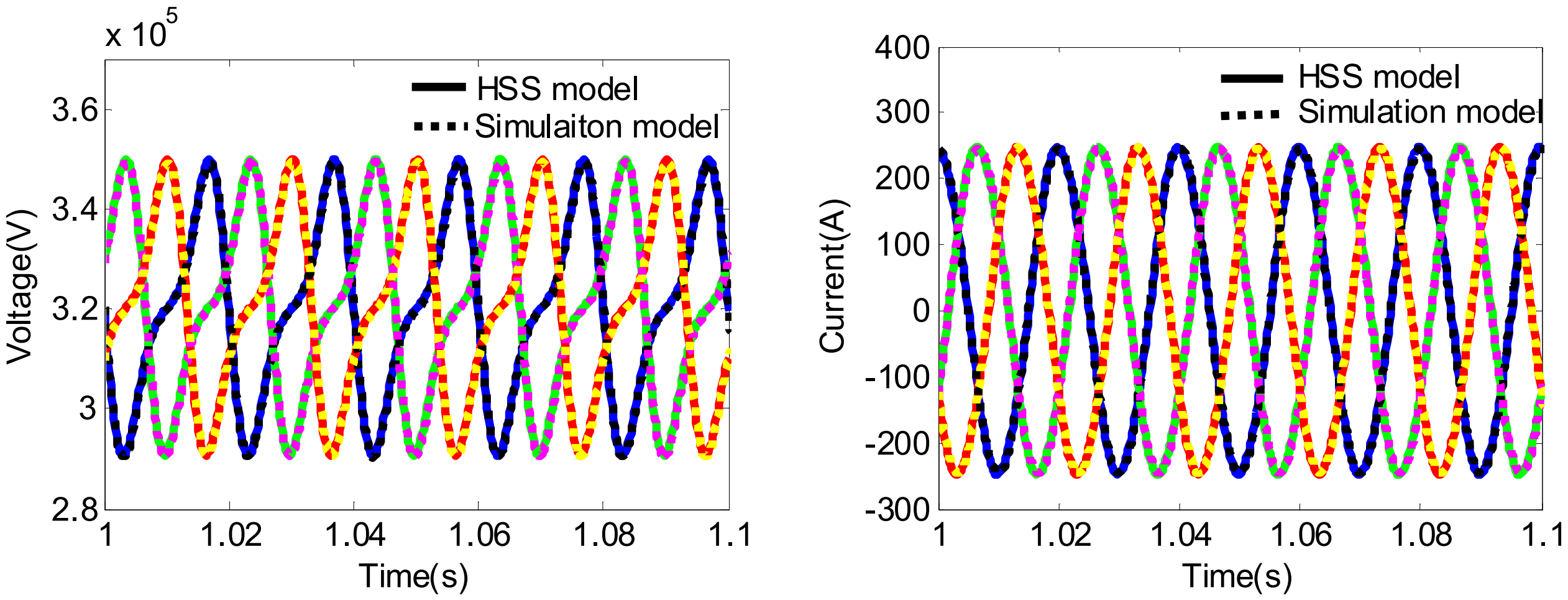

(c) (d)

Fig. 3. Validation for the steady state HSS model of the three-phase MMC. (a) Three-phase circulating currents. (b) Three-phase upper arm capacitor voltages. (c) Three-phase lower arm capacitor voltages. (d) Three-phase ac phase currents.

## IV. Small-Signal Dynamic HSS Model of a Three-Phase MMC

Taking the ac voltage closed-loop control for example in this paper, the small-signal dynamic HSS model of a three-phase MMC is developed. Fig. 4 depicts the block diagram of the ac voltage closed-loop control in the three-phase stationary frame, where $H_v(s)$ is a proportional-resonant (PR) controller to achieve zero steady-state errors for sinusoidal quantities, and $k_f$ is the feedforward gain to improve dynamic response.

The ac voltage regulator is

$$H_v\left(s\right)=K_p+\frac{K_r s}{s^2+\omega_1^2} \tag{25}$$

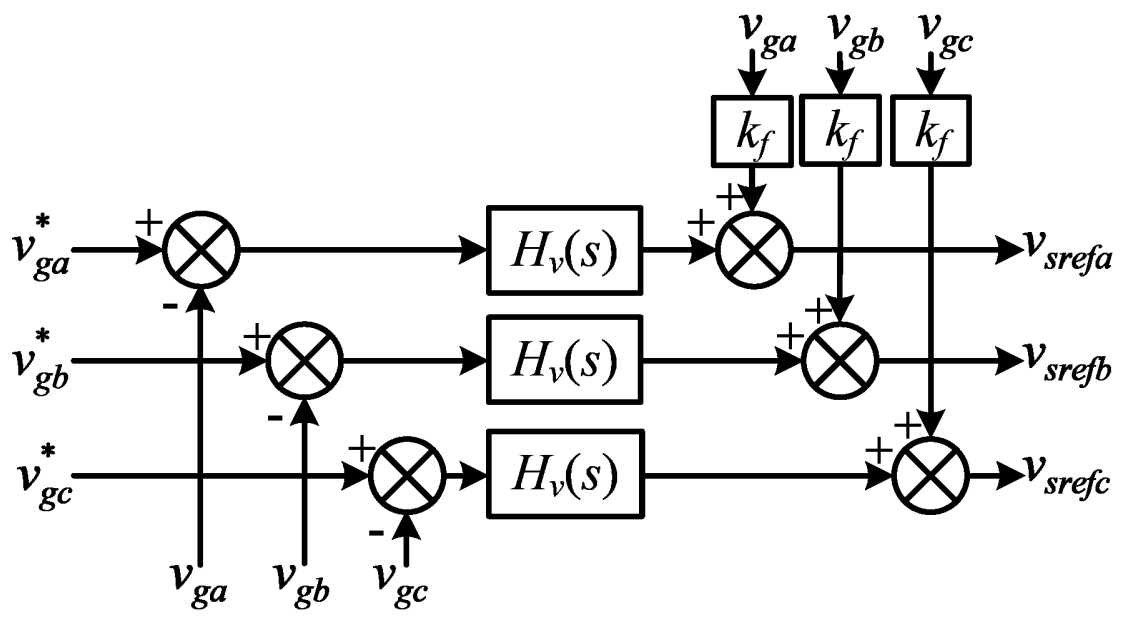

Fig. 4. Block diagram of the ac voltage closed-loop control.

Applying linearization to (10)~(14), we can obtain

$$\frac{d\Delta i_{cx}}{dt}=-\frac{R}{L}\Delta i_{cx}-\frac{n_{ux0}}{2L}\Delta v_{cux}^{\Sigma}-\frac{v_{cuxs}^{\Sigma}}{2L}\Delta n_{ux}-\frac{n_{lx0}}{2L}\Delta v_{clx}^{\Sigma}-\frac{v_{clxs}^{\Sigma}}{2L}\Delta n_{lx}+\frac{1}{2L}\Delta v_{dc},\left(x=a,b,c\right) \tag{26}$$

$$\frac{d\Delta v_{cux}^{\Sigma}}{dt}=\frac{n_{ux0}}{C_{arm}}\Delta i_{cx}+\frac{n_{ux0}}{2C_{arm}}\Delta i_{gx}+\frac{1}{C_{arm}}\left(i_{cxs}+\frac{i_{gxs}}{2}\right)\Delta n_{ux},\left(x=a,b,c\right) \tag{27}$$

$$\frac{d\Delta v_{clx}^{\Sigma}}{dt}=\frac{n_{lx0}}{C_{arm}}\Delta i_{cx}-\frac{n_{lx0}}{2C_{arm}}\Delta i_{gx}+\frac{1}{C_{arm}}\left(i_{cxs}-\frac{i_{gxs}}{2}\right)\Delta n_{lx},\left(x=a,b,c\right) \tag{28}$$

$$\frac{d\Delta i_{gx}}{dt} = -\frac{n_{ux0}}{L}\Delta v_{cux}^{\Sigma} - \frac{v_{cuxs}^{\Sigma}}{L}\Delta n_{ux} + \frac{n_{lx0}}{L}\Delta v_{clx}^{\Sigma} + \frac{v_{clxs}^{\Sigma}}{L}\Delta n_{lx} - \frac{R+2Z_L}{L}\Delta i_{gx}, (x = a,b,c) \tag{29}$$

where the subscript "0" and "$s$" in the insertion indices and state variables represent the steady state components, and the symbol "$\Delta$" means the small-signal components.

Additionally, the small-signal differential equations of the controller used in the MMC are

$$\frac{d\Delta x_{PRx1}}{dt} = -\omega_1^2 \Delta x_{PRx2} + K_r \Delta v_{gx}^{*} - K_r Z_L \Delta i_{gx}, (x = a,b,c) \tag{30}$$

$$\frac{d\Delta x_{PRx2}}{dt} = \Delta x_{PRx1}, (x = a,b,c) \tag{31}$$

$$\Delta n_{ux} = -\frac{1}{V_{dc}}\Delta x_{PRx1} - \frac{K_p}{V_{dc}}\Delta v_{gx}^{*} + \frac{K_p - K_f}{V_{dc}} Z_L \Delta i_{gx}, (x = a,b,c) \tag{32}$$

$$\Delta n_{lx} = \frac{1}{V_{dc}}\Delta x_{PRx1} + \frac{K_p}{V_{dc}}\Delta v_{gx}^{*} - \frac{K_p - K_f}{V_{dc}} Z_L \Delta i_{gx}, (x = a,b,c) \tag{33}$$

where $\Delta x_{PRx1}$ and $\Delta x_{PRx2}$ are the state variables used in the PR controller in the phase-$x$.

Hence, the state space form of the small-signal model of a three-phase MMC with ac voltage closed-loop control can written as

$$\Delta \dot{x}(t) = A(t)\Delta x(t) + B(t)\Delta u(t) \tag{34}$$

where

$$\begin{aligned} \Delta x(t) = \big[ & \Delta i_{ca}, \Delta i_{cb}, \Delta i_{cc}, \Delta v_{cua}^{\Sigma}, \Delta v_{cub}^{\Sigma}, \Delta v_{cuc}^{\Sigma}, \Delta v_{cla}^{\Sigma}, \Delta v_{clb}^{\Sigma}, \Delta v_{clc}^{\Sigma}, \Delta i_{ga}, \Delta i_{gb}, \Delta i_{gc}, \\ & \Delta x_{PRa1}, \Delta x_{PRa2}, \Delta x_{PRb1}, \Delta x_{PRb2}, \Delta x_{PRc1}, \Delta x_{PRc2} \big]^T \end{aligned} \tag{35}$$

$$\Delta u(t) = \left[ \Delta v_{dc}, \Delta v_{ga}^{*}, \Delta v_{gb}^{*}, \Delta v_{gc}^{*} \right]^T \tag{36}$$

$$A(t) = \begin{bmatrix}
-\frac{R}{L} & 0 & 0 & -\frac{n_{ua0}}{2L} & 0 & 0 & -\frac{n_{la0}}{2L} & 0 & 0 & F_{c1a} & 0 & 0 & F_{c2a} & 0 & 0 & 0 & 0 & 0 \\
0 & -\frac{R}{L} & 0 & 0 & -\frac{n_{ub0}}{2L} & 0 & 0 & -\frac{n_{lb0}}{2L} & 0 & 0 & F_{c1b} & 0 & 0 & 0 & F_{c2b} & 0 & 0 & 0 \\
0 & 0 & -\frac{R}{L} & 0 & 0 & -\frac{n_{uc0}}{2L} & 0 & 0 & -\frac{n_{lc0}}{2L} & 0 & 0 & F_{c1c} & 0 & 0 & 0 & 0 & F_{c2c} & 0 \\
\frac{n_{ua0}}{C_{arm}} & 0 & 0 & 0 & 0 & 0 & 0 & 0 & 0 & F_{vu1a} & 0 & 0 & F_{vu2a} & 0 & 0 & 0 & 0 & 0 \\
0 & \frac{n_{ub0}}{C_{arm}} & 0 & 0 & 0 & 0 & 0 & 0 & 0 & 0 & F_{vu1b} & 0 & 0 & 0 & F_{vu2b} & 0 & 0 & 0 \\
0 & 0 & \frac{n_{uc0}}{C_{arm}} & 0 & 0 & 0 & 0 & 0 & 0 & 0 & 0 & F_{vu1c} & 0 & 0 & 0 & 0 & F_{vu2c} & 0 \\
\frac{n_{la0}}{C_{arm}} & 0 & 0 & 0 & 0 & 0 & 0 & 0 & 0 & F_{vl1a} & 0 & 0 & F_{vl2a} & 0 & 0 & 0 & 0 & 0 \\
0 & \frac{n_{lb0}}{C_{arm}} & 0 & 0 & 0 & 0 & 0 & 0 & 0 & 0 & F_{vl1b} & 0 & 0 & 0 & F_{vl2b} & 0 & 0 & 0 \\
0 & 0 & \frac{n_{lc0}}{C_{arm}} & 0 & 0 & 0 & 0 & 0 & 0 & 0 & 0 & F_{vl1c} & 0 & 0 & 0 & 0 & F_{vl2c} & 0 \\
0 & 0 & 0 & -\frac{n_{ua0}}{L} & 0 & 0 & \frac{n_{la0}}{L} & 0 & 0 & F_{i1a} & 0 & 0 & F_{i2a} & 0 & 0 & 0 & 0 & 0 \\
0 & 0 & 0 & 0 & -\frac{n_{ub0}}{L} & 0 & 0 & \frac{n_{lb0}}{L} & 0 & 0 & F_{i1b} & 0 & 0 & 0 & F_{i2b} & 0 & 0 & 0 \\
0 & 0 & 0 & 0 & 0 & -\frac{n_{uc0}}{L} & 0 & 0 & \frac{n_{lc0}}{L} & 0 & 0 & F_{i1c} & 0 & 0 & 0 & 0 & F_{i2c} & 0 \\
0 & 0 & 0 & 0 & 0 & 0 & 0 & 0 & 0 & -K_r Z_L & 0 & 0 & 0 & -\omega_1^2 & 0 & 0 & 0 & 0 \\
0 & 0 & 0 & 0 & 0 & 0 & 0 & 0 & 0 & 0 & 0 & 0 & 1 & 0 & 0 & 0 & 0 & 0 \\
0 & 0 & 0 & 0 & 0 & 0 & 0 & 0 & 0 & 0 & -K_r Z_L & 0 & 0 & 0 & 0 & -\omega_1^2 & 0 & 0 \\
0 & 0 & 0 & 0 & 0 & 0 & 0 & 0 & 0 & 0 & 0 & 0 & 0 & 0 & 1 & 0 & 0 & 0 \\
0 & 0 & 0 & 0 & 0 & 0 & 0 & 0 & 0 & 0 & 0 & -K_r Z_L & 0 & 0 & 0 & 0 & 0 & -\omega_1^2 \\
0 & 0 & 0 & 0 & 0 & 0 & 0 & 0 & 0 & 0 & 0 & 0 & 0 & 0 & 0 & 0 & 1 & 0
\end{bmatrix} \tag{37}$$

$$B(t)=\begin{bmatrix} \frac{1}{2L} & F_{c3a} & 0 & 0 \\ \frac{1}{2L} & 0 & F_{c3b} & 0 \\ \frac{1}{2L} & 0 & 0 & F_{c3c} \\ 0 & F_{vu3a} & 0 & 0 \\ 0 & 0 & F_{vu3b} & 0 \\ 0 & 0 & 0 & F_{vu3c} \\ 0 & F_{vl3a} & 0 & 0 \\ 0 & 0 & F_{vl3b} & 0 \\ 0 & 0 & 0 & F_{vl3c} \\ 0 & F_{i3a} & 0 & 0 \\ 0 & 0 & F_{i3b} & 0 \\ 0 & 0 & 0 & F_{i3c} \\ 0 & K_r & 0 & 0 \\ 0 & 0 & 0 & 0 \\ 0 & 0 & K_r & 0 \\ 0 & 0 & 0 & 0 \\ 0 & 0 & 0 & K_r \\ 0 & 0 & 0 & 0 \end{bmatrix} \tag{38}$$

in which

$$F_{c1x}=\frac{\left(K_f-K_p\right)\left(v_{cuxs}^{\Sigma}-v_{clxs}^{\Sigma}\right)}{2LV_{dc}}Z_L,\ F_{c2x}=\frac{v_{cuxs}^{\Sigma}-v_{clxs}^{\Sigma}}{2LV_{dc}},\ F_{c3x}=\frac{K_p\left(v_{cuxs}^{\Sigma}-v_{clxs}^{\Sigma}\right)}{2LV_{dc}}$$

$$F_{vu1x}=\frac{n_{ux0}}{2C_{arm}}-\frac{\left(K_f-K_p\right)}{C_{arm}V_{dc}}\left(i_{cxs}+\frac{i_{gxs}}{2}\right)Z_L,\ F_{vu2x}=-\frac{1}{C_{arm}V_{dc}}\left(i_{cxs}+\frac{i_{gxs}}{2}\right),\ F_{vu3x}=-\frac{K_p}{C_{arm}V_{dc}}\left(i_{cxs}+\frac{i_{gxs}}{2}\right)$$

$$F_{vl1x}=\frac{n_{lx0}}{2C_{arm}}+\frac{\left(K_f-K_p\right)}{C_{arm}V_{dc}}\left(i_{cxs}-\frac{i_{gxs}}{2}\right)Z_L,\ F_{vl2x}=\frac{1}{C_{arm}V_{dc}}\left(i_{cxs}-\frac{i_{gxs}}{2}\right),\ F_{vl3x}=\frac{K_p}{C_{arm}V_{dc}}\left(i_{cxs}-\frac{i_{gxs}}{2}\right)$$

$$F_{i1x}=-\frac{R+2R_L}{L}+\frac{\left(K_f-K_p\right)\left(v_{cuxs}^{\Sigma}+v_{clxs}^{\Sigma}\right)}{LV_{dc}}Z_L,\ F_{i2x}=\frac{v_{cuxs}^{\Sigma}+v_{clxs}^{\Sigma}}{LV_{dc}},\ F_{i3x}=\frac{K_p\left(v_{cuxs}^{\Sigma}+v_{clxs}^{\Sigma}\right)}{LV_{dc}}$$

Then, by using the HSS modeling methodology, the small-signal dynamic HSS model of the three-phase MMC with ac voltage closed-loop control can be obtained as

$$s\Delta\mathbf{X}=\mathbf{A}\Delta\mathbf{X}+\mathbf{B}\Delta\mathbf{U} \tag{39}$$

where

$$\begin{aligned}\Delta\mathbf{X}=\big[&\Delta I_{ca},\Delta I_{cb},\Delta I_{cc},\Delta V_{cua}^{\Sigma},\Delta V_{cub}^{\Sigma},\Delta V_{cuc}^{\Sigma},\Delta V_{cla}^{\Sigma},\Delta V_{clb}^{\Sigma},\Delta V_{clc}^{\Sigma},\Delta I_{ga},\Delta I_{gb},\Delta I_{gc},\\ &\Delta X_{PRa1},\Delta X_{PRa2},\Delta X_{PRb1},\Delta X_{PRb2},\Delta X_{PRc1},\Delta X_{PRc2}\big]^T\end{aligned} \tag{40}$$

$$\Delta\mathbf{U}=\left[\Delta V_{dc},\Delta V_{ga}^{*},\Delta V_{gb}^{*},\Delta V_{gc}^{*}\right]^T \tag{41}$$

$$
\mathbf{A}=\begin{bmatrix}
-\frac{R}{L}I-Q & O & O & -\frac{\Gamma[N_{ua0}]}{2L} & O & O & -\frac{\Gamma[N_{la0}]}{2L} & O & O & \Gamma[F_{c1a}] & O & O & \Gamma[F_{c2a}] & O & O & O & O & O \\
O & -\frac{R}{L}I-Q & O & O & -\frac{\Gamma[N_{ub0}]}{2L} & O & O & -\frac{\Gamma[N_{lb0}]}{2L} & O & O & \Gamma[F_{c1b}] & O & O & O & \Gamma[F_{c2b}] & O & O & O \\
O & O & -\frac{R}{L}I-Q & O & O & -\frac{\Gamma[N_{uc0}]}{2L} & O & O & -\frac{\Gamma[N_{lc0}]}{2L} & O & O & \Gamma[F_{c1c}] & O & O & O & O & \Gamma[F_{c2c}] & O \\
\frac{\Gamma[N_{ua0}]}{C_{arm}} & O & O & -Q & O & O & O & O & O & \Gamma[F_{vu1a}] & O & O & \Gamma[F_{vu2a}] & O & O & O & O & O \\
O & \frac{\Gamma[N_{ub0}]}{C_{arm}} & O & O & -Q & O & O & O & O & O & \Gamma[F_{vu1b}] & O & O & O & \Gamma[F_{vu2b}] & O & O & O \\
O & O & \frac{\Gamma[N_{uc0}]}{C_{arm}} & O & O & -Q & O & O & O & O & O & \Gamma[F_{vu1c}] & O & O & O & O & \Gamma[F_{vu2c}] & O \\
\frac{\Gamma[N_{la0}]}{C_{arm}} & O & O & O & O & O & -Q & O & O & \Gamma[F_{vl1a}] & O & O & \Gamma[F_{vl2a}] & O & O & O & O & O \\
O & \frac{\Gamma[N_{lb0}]}{C_{arm}} & O & O & O & O & O & -Q & O & O & \Gamma[F_{vl1b}] & O & O & O & \Gamma[F_{vl2b}] & O & O & O \\
O & O & \frac{\Gamma[N_{lc0}]}{C_{arm}} & O & O & O & O & O & -Q & O & O & \Gamma[F_{vl1c}] & O & O & O & O & \Gamma[F_{vl2c}] & O \\
O & O & O & -\frac{\Gamma[N_{ua0}]}{L} & O & O & \frac{\Gamma[N_{la0}]}{L} & O & O & \Gamma[F_{i1a}]-Q & O & O & \Gamma[F_{i2a}] & O & O & O & O & O \\
O & O & O & O & -\frac{\Gamma[N_{ub0}]}{L} & O & O & \frac{\Gamma[N_{lb0}]}{L} & O & O & \Gamma[F_{i1b}]-Q & O & O & O & \Gamma[F_{i2b}] & O & O & O \\
O & O & O & O & O & -\frac{\Gamma[N_{uc0}]}{L} & O & O & \frac{\Gamma[N_{lc0}]}{L} & O & O & \Gamma[F_{i1c}]-Q & O & O & O & O & \Gamma[F_{i2c}] & O \\
O & O & O & O & O & O & O & O & O & -K_r Z_L I & O & O & -Q & -\omega_1^2 I & O & O & O & O \\
O & O & O & O & O & O & O & O & O & O & O & O & I & -Q & O & O & O & O \\
O & O & O & O & O & O & O & O & O & O & -K_r Z_L I & O & O & O & -Q & -\omega_1^2 I & O & O \\
O & O & O & O & O & O & O & O & O & O & O & O & O & O & I & -Q & O & O \\
O & O & O & O & O & O & O & O & O & O & O & -K_r Z_L I & O & O & O & O & -Q & -\omega_1^2 I \\
O & O & O & O & O & O & O & O & O & O & O & O & O & O & O & O & I & -Q
\end{bmatrix} \quad (42)
$$

$$
\mathbf{B}=\begin{bmatrix}
\frac{1}{2L}I & \Gamma[F_{c3a}] & O & O \\
\frac{1}{2L}I & O & \Gamma[F_{c3b}] & O \\
\frac{1}{2L}I & O & O & \Gamma[F_{c3c}] \\
O & \Gamma[F_{vu3a}] & O & O \\
O & O & \Gamma[F_{vu3b}] & O \\
O & O & O & \Gamma[F_{vu3c}] \\
O & \Gamma[F_{vl3a}] & O & O \\
O & O & \Gamma[F_{vl3b}] & O \\
O & O & O & \Gamma[F_{vl3c}] \\
O & \Gamma[F_{i3a}] & O & O \\
O & O & \Gamma[F_{i3b}] & O \\
O & O & O & \Gamma[F_{i3c}] \\
O & K_r I & O & O \\
O & O & O & O \\
O & O & K_r I & O \\
O & O & O & O \\
O & O & O & K_r I \\
O & O & O & O
\end{bmatrix} \quad (43)
$$

in which, the Toeplitz matrices $\Gamma\left[V_{cuxs}^{\Sigma}-V_{clxs}^{\Sigma}\right]$, $\Gamma\left[V_{cuxs}^{\Sigma}+V_{clxs}^{\Sigma}\right]$, $\Gamma\left[I_{cxs}+\frac{I_{gxs}}{2}\right]$, and $\Gamma\left[I_{cxs}-\frac{I_{gxs}}{2}\right]$ need to be calculated, where the steady state components of the harmonic coefficients of the state variables can be calculated by (24).

In order to verify the developed small-signal dynamic HSS model of the three-phase MMC with ac voltage closed-loop control, a comparison between the results from the nonlinear time domain simulation model and the small-signal dynamic HSS model has been carried out, as shown in Fig. 5, where a step change with 10 kV fundamental voltage in the phase-a reference voltage is made at 1.5 s. It can be seen that both the phase-a circulating current and ac phase current become larger after a step change in phase-a reference voltage, and the results from the small-signal dynamic HSS model of MMC have a good match with those from the nonlinear time domain simulation model.

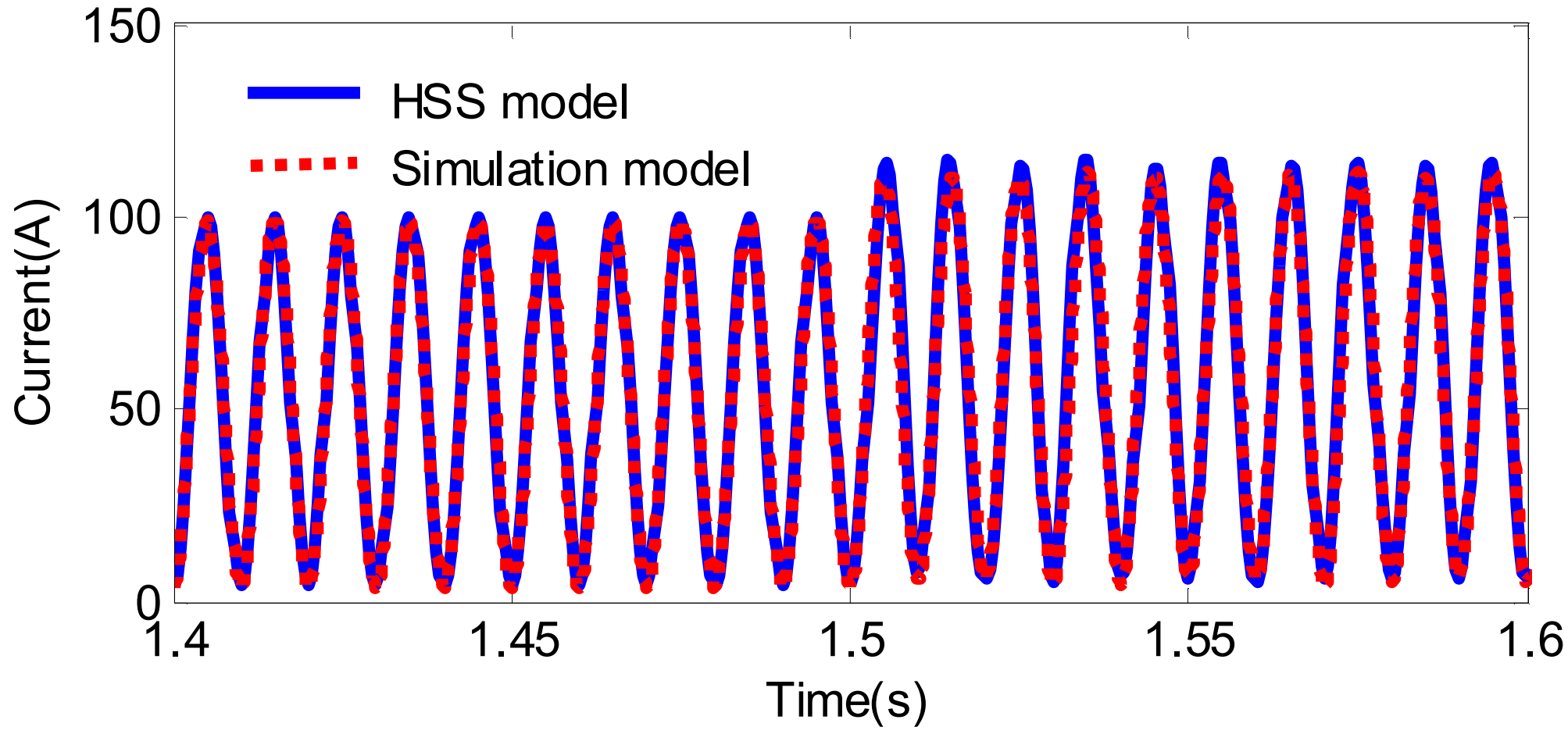

(a)

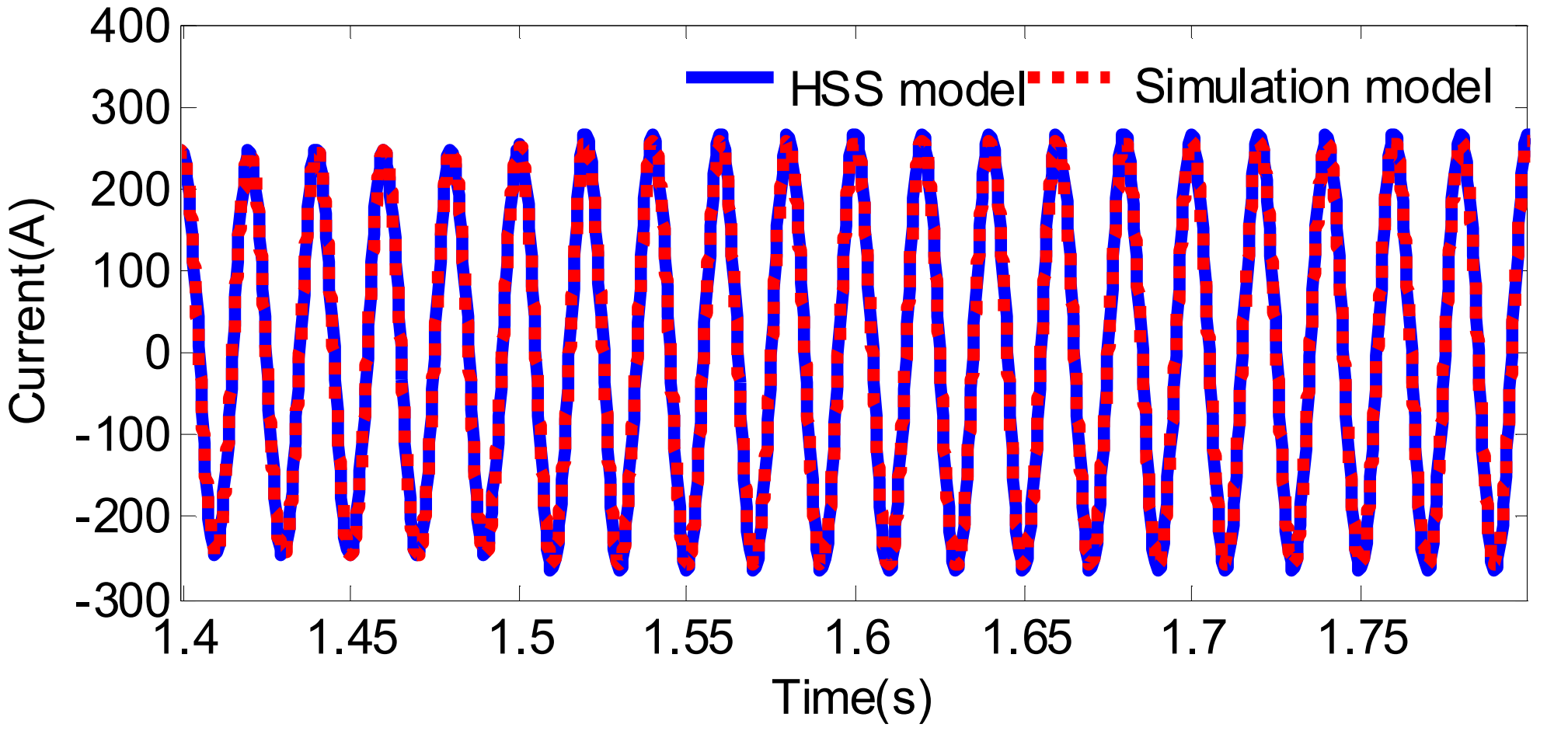

(b)

Fig. 5. Validation for the small-signal dynamic HSS model of the three-phase MMC. (a) Phase-a circulating current. (b) Phase-a ac phase current.

## V. EXPERIMENTAL RESULTS

To further validate the HSS model of a three-phase MMC developed in this paper, the experimental measurements on a three-phase MMC laboratory setup have also been carried out. The topology of the MMC setup is identical to that in Fig. 1. A dc power source is connected to the dc terminals of the MMC, and the three-phase resistive load is connected to the ac terminals of the MMC. The parameters of the MMC prototype are listed in Table I. In this experiment, the dc voltage of 450 V and ac phase voltage amplitude of 200 V are used. In addition, the load resistance $R_L = 10\ \Omega$. Fig. 6 shows the photograph of the MMC laboratory setup, where only phase-a can be seen in this view, and phase-b and phase-c are in the back of this setup. The MMC setup is controlled by the digital controller which is based on OMAP-L137 (C6747 DSP + ARM926) and Spartan-3A FPGA. LabView is used as the human interface where program loading, online parameter tuning, signal monitoring, and start-stop control can be manipulated.

TABLE I. PARAMETERS OF MMC PROTOTYPE

| **Parameter** | **Value** |
|---|---|
| Rated power | 30 kW |
| Rated frequency | 50 Hz |
| Rated dc voltage | 1000 V |
| Rated ac voltage | 520 V |
| Submodule number per arm | 12 |
| Submodule capacitor | 6.6 mF |
| Arm inductance | 5 mH |
| Switching frequency | 762 Hz |

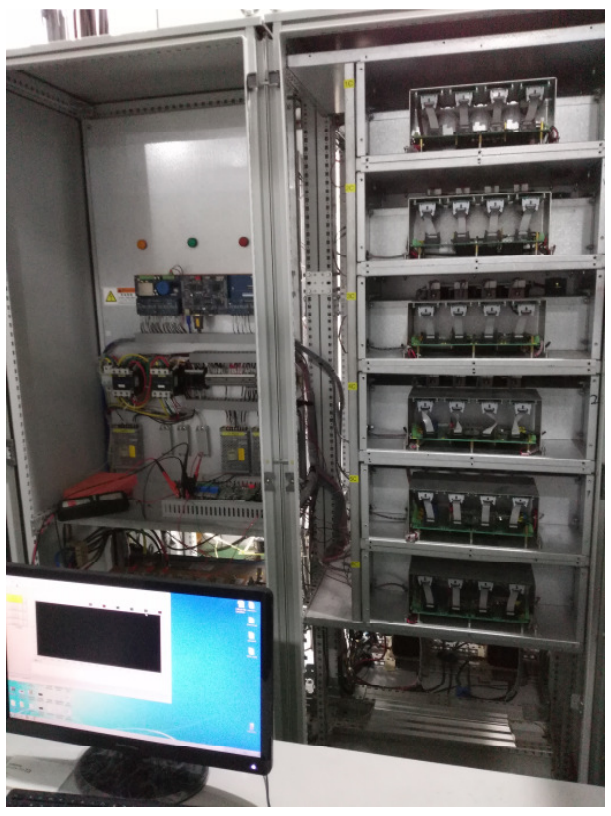

Fig. 6. Photograph of the laboratory setup.

Open-loop control is used in this experiment to validate the steady state HSS model of the three-phase MMC. Fig. 7 presents the results from the HSS model and experimental setup, where the three-phase upper arm capacitor voltages, three-phase lower arm capacitor voltages, and three-phase ac phase currents are demonstrated, respectively. The solid lines represent the results from the HSS model, and the dashed lines denote the results from the experimental setup. It can be observed that the results from the HSS model and experimental setup are in substantial agreement. However, small phase differences exist between the two results, which are mainly caused by the modulation link (i.e. PWM). Due to the relatively low switching frequency (762 Hz), the PWM delay (normally half-carrier period) becomes relatively large. In this experiment, the theoretical value of the PWM delay is 656.2 μs that corresponds to the phase difference of 11.8 °. From Fig. 7(c), a phase difference of approximately 12° between the two currents can be measured, which is consistent with the theoretical value. Furthermore, the small discrepancies in amplitude seen may be the results of actual parameter variations (e.g. asymmetric circuit parameters), nonlinear characteristics (e.g. dead-time effect), sampling channel difference, and so on.

The experimental results for the validation of the small-signal dynamic HSS model of a three-phase MMC will be added later. In addition, the delay link will also be considered in the developed HSS model later.

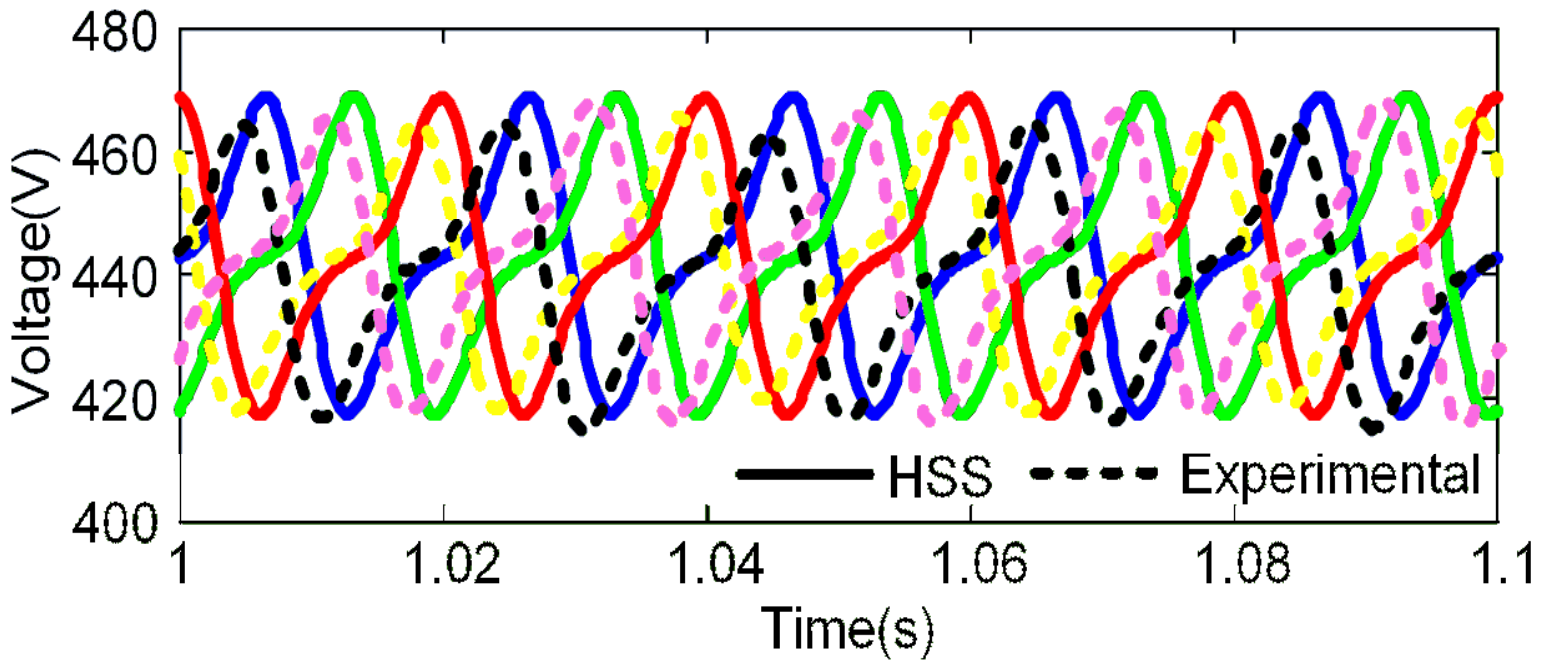

(a)

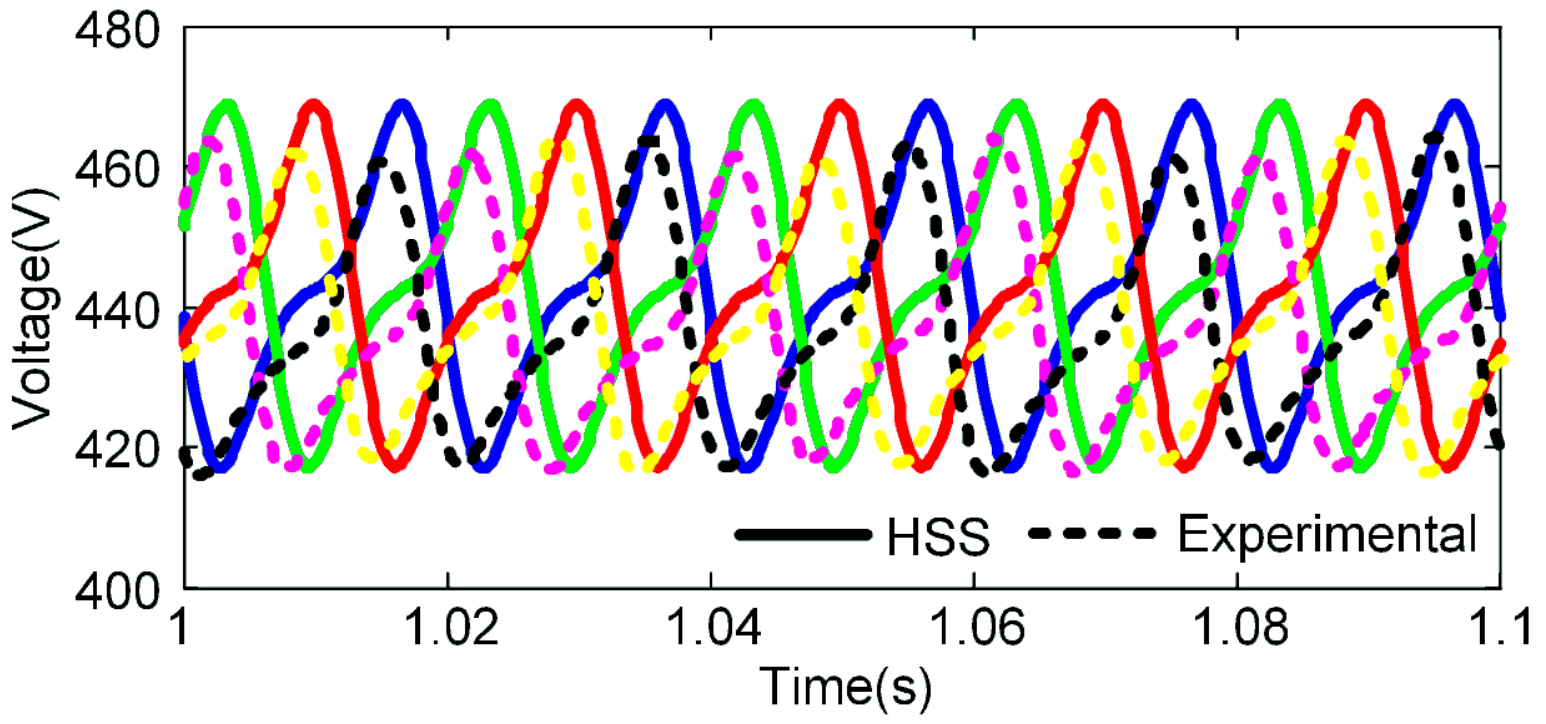

(b)

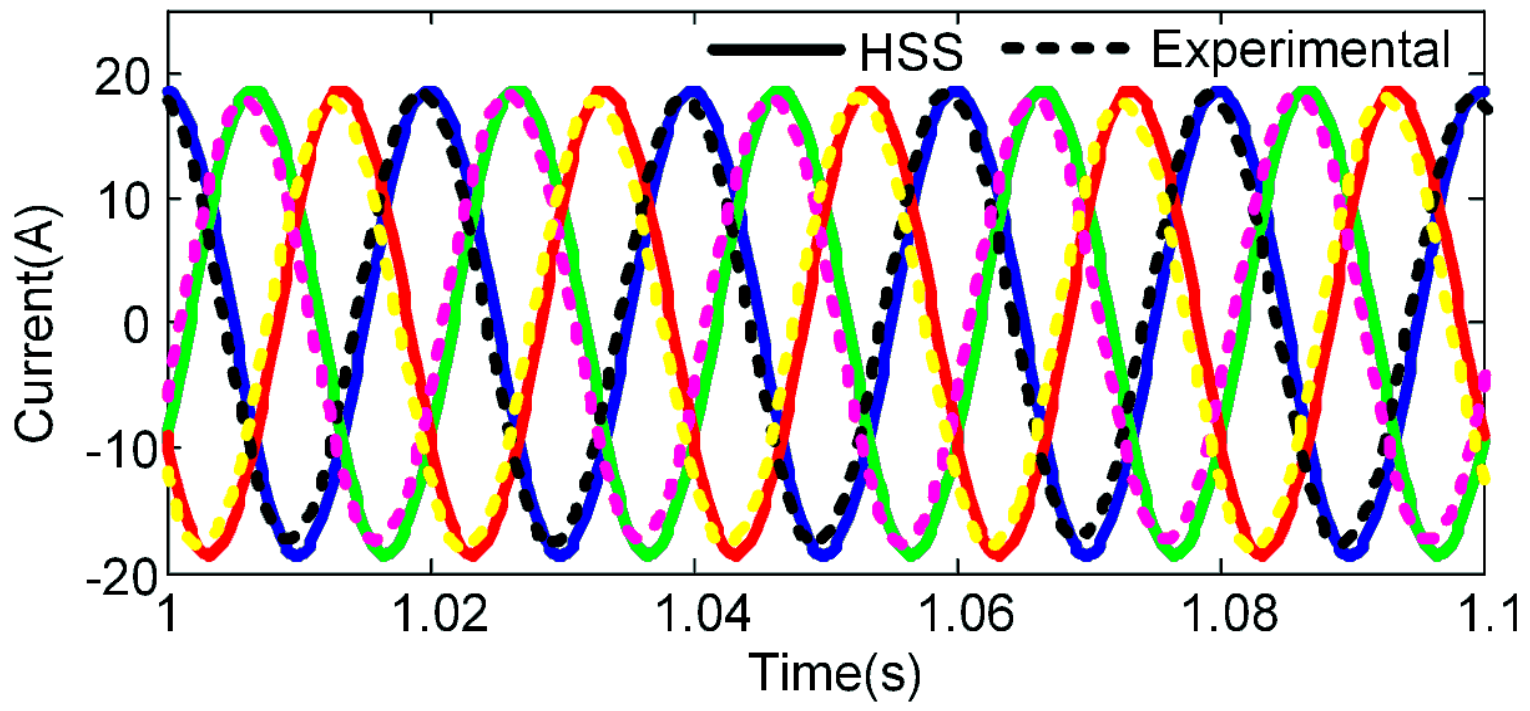

(c)

Fig. 7. Validation for the steady state HSS model of the three-phase MMC. (a) Three-phase upper arm capacitor voltages. (b) Three-phase lower arm capacitor voltages. (c) Three-phase ac phase currents.

## VI. Conclusion

This paper presents the harmonic state space (HSS) modeling of a three-phase MMC. The steady state and small-signal dynamic HSS models of a three-phase MMC have been developed in this paper, respectively. The validity of the developed HSS models of the three-phase MMC has been verified by both the nonlinear time domain simulation model in MATLAB/Simulink and the experimental setup. The results show that the developed HSS models of a three-phase MMC are able to capture all the harmonics in the capacitor voltages and circulating currents, which are accurate enough for steady state harmonic studies and for the dynamic small-signal studies. Furthermore, the HSS models of a three-phase MMC developed in this paper are easily extended to the high number of harmonics, which just increase the harmonic order in the matrices of the HSS model.

## REFERENCES

The Toeplitz matrices of the Fourier coefficients of the insertion indices are given in (A1)~(A6).

$$\Gamma[N_{ua0}]=\begin{bmatrix} \frac{1}{2} & -\frac{m}{4} & 0 & 0 & 0 & 0 & 0 \\ -\frac{m}{4} & \frac{1}{2} & -\frac{m}{4} & 0 & 0 & 0 & 0 \\ 0 & -\frac{m}{4} & \frac{1}{2} & -\frac{m}{4} & 0 & 0 & 0 \\ 0 & 0 & -\frac{m}{4} & \frac{1}{2} & -\frac{m}{4} & 0 & 0 \\ 0 & 0 & 0 & -\frac{m}{4} & \frac{1}{2} & -\frac{m}{4} & 0 \\ 0 & 0 & 0 & 0 & -\frac{m}{4} & \frac{1}{2} & -\frac{m}{4} \\ 0 & 0 & 0 & 0 & 0 & -\frac{m}{4} & \frac{1}{2} \end{bmatrix} \tag{A1}$$

$$\Gamma[N_{ub0}]=\begin{bmatrix} \frac{1}{2} & \frac{m(1-j\sqrt{3})}{8} & 0 & 0 & 0 & 0 & 0 \\ \frac{m(1+j\sqrt{3})}{8} & \frac{1}{2} & \frac{m(1-j\sqrt{3})}{8} & 0 & 0 & 0 & 0 \\ 0 & \frac{m(1+j\sqrt{3})}{8} & \frac{1}{2} & \frac{m(1-j\sqrt{3})}{8} & 0 & 0 & 0 \\ 0 & 0 & \frac{m(1+j\sqrt{3})}{8} & \frac{1}{2} & \frac{m(1-j\sqrt{3})}{8} & 0 & 0 \\ 0 & 0 & 0 & \frac{m(1+j\sqrt{3})}{8} & \frac{1}{2} & \frac{m(1-j\sqrt{3})}{8} & 0 \\ 0 & 0 & 0 & 0 & \frac{m(1+j\sqrt{3})}{8} & \frac{1}{2} & \frac{m(1-j\sqrt{3})}{8} \\ 0 & 0 & 0 & 0 & 0 & \frac{m(1+j\sqrt{3})}{8} & \frac{1}{2} \end{bmatrix} \tag{A2}$$

$$\Gamma[N_{uc0}]=\begin{bmatrix} \frac{1}{2} & \frac{m(1+j\sqrt{3})}{8} & 0 & 0 & 0 & 0 & 0 \\ \frac{m(1-j\sqrt{3})}{8} & \frac{1}{2} & \frac{m(1+j\sqrt{3})}{8} & 0 & 0 & 0 & 0 \\ 0 & \frac{m(1-j\sqrt{3})}{8} & \frac{1}{2} & \frac{m(1+j\sqrt{3})}{8} & 0 & 0 & 0 \\ 0 & 0 & \frac{m(1-j\sqrt{3})}{8} & \frac{1}{2} & \frac{m(1+j\sqrt{3})}{8} & 0 & 0 \\ 0 & 0 & 0 & \frac{m(1-j\sqrt{3})}{8} & \frac{1}{2} & \frac{m(1+j\sqrt{3})}{8} & 0 \\ 0 & 0 & 0 & 0 & \frac{m(1-j\sqrt{3})}{8} & \frac{1}{2} & \frac{m(1+j\sqrt{3})}{8} \\ 0 & 0 & 0 & 0 & 0 & \frac{m(1-j\sqrt{3})}{8} & \frac{1}{2} \end{bmatrix} \tag{A3}$$

$$\Gamma[N_{la0}] = \begin{bmatrix} \frac{1}{2} & \frac{m}{4} & 0 & 0 & 0 & 0 & 0 \\ \frac{m}{4} & \frac{1}{2} & \frac{m}{4} & 0 & 0 & 0 & 0 \\ 0 & \frac{m}{4} & \frac{1}{2} & \frac{m}{4} & 0 & 0 & 0 \\ 0 & 0 & \frac{m}{4} & \frac{1}{2} & \frac{m}{4} & 0 & 0 \\ 0 & 0 & 0 & \frac{m}{4} & \frac{1}{2} & \frac{m}{4} & 0 \\ 0 & 0 & 0 & 0 & \frac{m}{4} & \frac{1}{2} & \frac{m}{4} \\ 0 & 0 & 0 & 0 & 0 & \frac{m}{4} & \frac{1}{2} \end{bmatrix} \tag{A4}$$

$$\Gamma[N_{lb0}] = \begin{bmatrix} \frac{1}{2} & \frac{m\left(-1+j\sqrt{3}\right)}{8} & 0 & 0 & 0 & 0 & 0 \\ -\frac{m\left(1+j\sqrt{3}\right)}{8} & \frac{1}{2} & \frac{m\left(-1+j\sqrt{3}\right)}{8} & 0 & 0 & 0 & 0 \\ 0 & -\frac{m\left(1+j\sqrt{3}\right)}{8} & \frac{1}{2} & \frac{m\left(-1+j\sqrt{3}\right)}{8} & 0 & 0 & 0 \\ 0 & 0 & -\frac{m\left(1+j\sqrt{3}\right)}{8} & \frac{1}{2} & \frac{m\left(-1+j\sqrt{3}\right)}{8} & 0 & 0 \\ 0 & 0 & 0 & -\frac{m\left(1+j\sqrt{3}\right)}{8} & \frac{1}{2} & \frac{m\left(-1+j\sqrt{3}\right)}{8} & 0 \\ 0 & 0 & 0 & 0 & -\frac{m\left(1+j\sqrt{3}\right)}{8} & \frac{1}{2} & \frac{m\left(-1+j\sqrt{3}\right)}{8} \\ 0 & 0 & 0 & 0 & 0 & -\frac{m\left(1+j\sqrt{3}\right)}{8} & \frac{1}{2} \end{bmatrix} \tag{A5}$$

$$\Gamma[N_{lc0}] = \begin{bmatrix} \frac{1}{2} & -\frac{m\left(1+j\sqrt{3}\right)}{8} & 0 & 0 & 0 & 0 & 0 \\ \frac{m\left(-1+j\sqrt{3}\right)}{8} & \frac{1}{2} & -\frac{m\left(1+j\sqrt{3}\right)}{8} & 0 & 0 & 0 & 0 \\ 0 & \frac{m\left(-1+j\sqrt{3}\right)}{8} & \frac{1}{2} & -\frac{m\left(1+j\sqrt{3}\right)}{8} & 0 & 0 & 0 \\ 0 & 0 & \frac{m\left(-1+j\sqrt{3}\right)}{8} & \frac{1}{2} & -\frac{m\left(1+j\sqrt{3}\right)}{8} & 0 & 0 \\ 0 & 0 & 0 & \frac{m\left(-1+j\sqrt{3}\right)}{8} & \frac{1}{2} & -\frac{m\left(1+j\sqrt{3}\right)}{8} & 0 \\ 0 & 0 & 0 & 0 & \frac{m\left(-1+j\sqrt{3}\right)}{8} & \frac{1}{2} & -\frac{m\left(1+j\sqrt{3}\right)}{8} \\ 0 & 0 & 0 & 0 & 0 & \frac{m\left(-1+j\sqrt{3}\right)}{8} & \frac{1}{2} \end{bmatrix} \tag{A6}$$